\documentclass[10pt,a4paper]{article}
\usepackage{cite,a4}
\advance\textheight by -\headsep\advance\textheight by -\headheight
\newcommand{\eqs}{Eqs.$\;$}

\newcommand{\abc}[1]{\mbox{#1)}\quad}

\newcommand{\sqr}[2]{\; \sqrt[\scriptstyle
\kern-.5em{\mbox{\small #1}}\ ]{#2}}
\newcommand{\dd}{\mbox{d}}
\begin{document}

\title{Generation of exact solutions in cosmology on the basis of
five-dimensional Projective Unified Field Theory}
\author{
A.A. Blinkouski{\thanks{E-mail: blinkouski@yandex.ru}}\\
International Sakharov Environmental University\\
23 Dolgobrodskaya St., Minsk 220009, Belarus \and A.K.
Gorbatsievich{\thanks {E-mail: gorbatsievich@bsu.by}}\\
Department of Theoretical Physics, Belarusian State University \\
4 Skaryna Av., Minsk 220050, Belarus}
\date{}
\maketitle \noindent {\bf Abstract.}\\[1ex] {\small A homogeneous and
isotropic model of the Universe is considered in the framework of the
five-dimensional Projective Unified Field Theory in which the
gravitation is described by both space-time curvature and some
hypothetical scalar field ($\sigma$-field). We propose a generation
method for obtaining exact solutions. New exact Friedmann-like
solutions for a dust model and inflationary solutions are found. It
is shown that in the framework of exponential type inflation we
obtain a natural explanation of why at present we do not observe
$\sigma$-field effects or why these effects are negligible.}\\[1ex]

\thispagestyle{empty}

\vspace{4ex}
\section{Introduction}

As is well known, the idea of a 5-dimensional unified field theory
goes back to the works of Kaluza and Klein \cite{Kaluza,Klein}.
The pioneers of the projective approach to this theory were Veblen
and van Dantzig~\cite{Veblen,Dantzig}. Later this approach was
further developed by many authors (see the corresponding
references and a review of other higher-dimensional unified
theories in~\cite{Vladimirov87,Vladimirov98,Overduin}).

In this paper, a homogeneous and isotropic model of the Universe
is considered in the framework of the 5-dimensional Projective
Unified Field Theory (PUFT) developed by E. Schmutzer
\cite{Schmutzer68, Schmutzer83, Schmutzer95}. In PUFT, gravitation
is described by both space-time curvature and some hypothetical
scalar field ($\sigma$-field). To characterize the scalar field
predicted in PUFT as a new fundamental phenomenon in Nature,
Schmutzer introduced the notion ``scalarism'' (adjective:
scalaric) by analogy with electromagnetism. A source of this
``scalaric'' field can be both the electromagnetic field and a new
attribute of matter named by E. Schmutzer scalaric mass.

The PUFT is based on the postulated 5-dimensional Einstein-like
field equations. By projecting them into the 4-dimensional
space-time  one can obtain the following 4-dimensional field
equations (the cosmological term is omitted here)
\cite{Schmutzer83}:
\begin{eqnarray} \label {2.1}
R_{mn}-\frac{1}{2}\;g_{mn}\;R
=\kappa_0\left(E_{mn}+\Sigma_{mn}+\Theta_{mn}\right)
\end{eqnarray}
are the generalized gravitational field equations;
\begin{eqnarray} \label {2.2}
\abc {a} H ^ {mn} {} _ {; n} = \frac {4 \pi} {c} j ^ {m}, \quad \abc
{b}B_{mn,k}+B_{km,n}+B_{nk,m} = 0, \quad \abc {c} H _ {mn} = e ^ {3 \sigma} B
_ {mn}
\end{eqnarray}
are the generalized electromagnetic field equations;
\begin{eqnarray} \label {2.3}
\sigma ^ {, k} {} _ {; k} = \kappa _ 0 \left
(\frac {2} {3} \vartheta + \frac {1} {8\pi} B _ {ik} H ^ {ik} \right)
\end{eqnarray}
is the scalaric field equation. Here $R_{mn}$ is the Ricci tensor,
\begin{eqnarray} \label {2.4}
E _ {mn} = \frac {1} {4\pi} \left (B
_ {mk} H ^ {k} {} _ {n} + \frac {1} {4} g _ {mn} B _ {ik} H ^ {ik} \right)
\end{eqnarray}
is the electromagnetic energy-momentum tensor,
\begin{eqnarray}\label {2.5}
\Sigma_{mn} = - \frac {3} {2\kappa_0}
\left (\sigma_{,m} \sigma_{,n} - \frac{1}{2} g_{mn}
\sigma _ {, k} \sigma ^ {, k} \right)
\end{eqnarray}
is the scalaric energy-momentum tensor, $\Theta_{mn}$ is the
energy-momentum tensor of the nongeometrized matter (substrate),
$H_{mn}$ and $B_{mn}$ are the electromagnetic induction and the
field strength tensor, respectively, $j^{k}$ is the electric
current density, $\vartheta $ is the scalaric substrate density,
$\kappa_{0}=8\pi G/c^{4}$ is Einstein's gravitational constant
($G$ is Newton's gravitational constant). Latin indices run from 1
to 4; the comma and semicolon denote  partial and covariant
derivatives, respectively; the signature of the metric is $+2$.

These field equations lead to the following generalized energy
conservation law  and continuity equation for electric current
density:
\begin {eqnarray} \label {2.6}
\abc {a} \Theta ^{mn}{}_{;n} = -\frac{1}{c}B^{m}{}_{k}
         j^{k} + \vartheta \sigma^{,m} ,   \qquad
\abc {b} j^{m}{}_{;m} = 0.
\end {eqnarray}
It should be noted that recently E. Schmutzer offered a new
variant of PUFT (see \cite{Schmutzer95} and references therein)
with  slightly different 4-dimensional field equations as compared
to the above-stated ones (one can find a detailed analysis of the
geometric axiomatics of PUFT in \cite{Gorbatsievich}).  Both
variants are physically acceptable and deserve a comprehensive
investigation. An analysis of \eqs (\ref{2.1})--(\ref{2.5}) shows
that all the subsequent reasonings can be easily extended to the
last version of PUFT.

\section{Cosmological equations and generation of exact solutions }
Let us consider a homogeneous and isotropic cosmological model
with the Robertson-Walker line element in the well-known form:
\begin{eqnarray} \label {3.1}
\dd s^{2}=R^2(t) \left[ \frac {\dd r^2}{1-kr^2} + r^2
(\dd \theta^2 + \sin^2 \theta \; \dd \varphi^2) \right]
- c^2 \dd t^2,
\end{eqnarray}
where $R(t)$ is the scale factor and $k$ takes the values $0$ or $
\pm 1 $. For an electrically neutral continuum, described by the
energy-momentum tensor of a perfect fluid, \eqs (\ref{2.1}),
(\ref{2.3}) in the metric (\ref{3.1})  lead to the following set
of equations (the dot denotes a time derivative, $\varrho$ is the
mass density, $p$  is the  pressure):
\begin{eqnarray} \label{3.2}
\frac{\ddot{R}}{R} = - \frac{\kappa_0 c^2}{6} \left( \varrho c^2  +
3p \right) - \frac{1}{2} \dot{\sigma}^2 ,
\end{eqnarray}
\begin{eqnarray} \label{3.3}
\frac{\ddot{R}}{R}  +  \frac{2(\dot{R}^2 + k c^2 )}{R^2} =
\frac{\kappa_0 c^2}{2} \left( \varrho c^2 - p \right),
\end{eqnarray}
\begin{eqnarray} \label{3.4}
\ddot{ \sigma } + 3 \frac{  \dot{R}  }{R} \; \dot{  \sigma  }  =
- \frac{2}{3}\; \kappa_0 c^2 \vartheta ,
\end{eqnarray}
while the generalized energy conservation law (\ref{2.6}) gives
\begin{eqnarray} \label{3.5}
\dot{\varrho} + 3 \frac{ \dot{R} }{R} \left(  \varrho  +  \frac{p}{c^2}
\right) = \frac{\vartheta}{c^2} \dot{\sigma} .
\end{eqnarray}
\eqs (\ref{3.2}) to (\ref{3.5}) determine the  dynamics of the
cosmological model if the  equations  of  state, i.e.
$p=p(\varrho)$ and $\vartheta=\vartheta(\varrho)$, are given. The
Friedmann model corresponds to the special case $\vartheta = 0$
and $\dot{\sigma}=0$ of our model. Unfortunately, the above set of
differential equations leads \cite{Schmutzer1a} to an Abel
differential equation  and till now was solved exactly only for
some special
cases~\cite{Schmutzer1a,Schmutzer2a,Herlt,Blinkouski1}.

To obtain new exact solutions, let us consider the problem in a
slightly different aspect. Firstly, we shall use the arbitrariness
in the choice of an equation of state
$\vartheta=\vartheta(\varrho)$, because the functional form
$\vartheta(\varrho)$ is not determined within the theory.
Secondly, notice that above equations are not independent. For
instance, it is possible to show that the last \eqs (\ref{3.5}) is
a differential consequence of (\ref{3.2})--(\ref{3.4}). Therefore
on the further treatment we shall use only \eqs
(\ref{3.2})--(\ref{3.4}). If the equation of state for ordinary
matter is specified, $p=\nu \varrho c^{2} \ (-1\leq \nu <1)$, then
\eqs (\ref{3.2})--(\ref{3.4}) can be reduced to a form where the
functions $\sigma(t)$, $\varrho(t)$ and $\vartheta(t)\equiv
\vartheta(\varrho(t))$ are expressed in terms of the function
$R(t)$ and its derivatives:
\begin{eqnarray} \label{3.6}
\sigma(t)= \pm \; \sqrt{
\frac{8}{3(1-\nu)}}
\int \sqrt{
-\;\frac{\ddot{R}}{R} - \frac{1+3\nu}{2} \; \frac{\dot{R}^2 + kc^2}{R^2}
}\;
\dd t + \sigma_{0}  \;,
\end{eqnarray}
\begin{eqnarray} \label{3.7}
\varrho(t)= \frac{2}{(1-\nu)\kappa_0 c^4} \left[
\frac{\ddot{R}}{R} +  \frac{2(\dot{R}^2+kc^2)}{R^2} \right],
\end{eqnarray}
\begin{eqnarray} \label{3.8}
\vartheta(t)= -\frac{3}{2\kappa_{0} c^{2} R^{3}}\;
\frac{\dd}{\dd t} \left( \dot{\sigma}R^{3} \right).
\end{eqnarray}
Here $\sigma_{0}$ is an integration constant and $\nu \not=1$ (it
is necessary to consider the case $\nu=1$  separately). If we
specify the time dependence of the scale factor, $R(t)$, we can
find the corresponding functions $\sigma(t)$, $\varrho(t)$ and
$\vartheta(t)$ which are necessary for this Universe evolution
scenario. However, the choice of the dependence $R(t)$ is not
free, because the natural requirement $\varrho \geq 0$, taking
into account (\ref{3.6}) and (\ref{3.7}), gives the following
restrictions on the choice of the scale factor:
\begin{eqnarray} \label{3.9}
-\;\frac{\ddot{R}}{R} - \frac{1+3\nu}{2} \; \frac{\dot{R}^2 + kc^2}{R^2}
\geq 0 ,
\end{eqnarray}
\begin{eqnarray} \label{3.10}
\frac{\ddot{R}}{R}  +  \frac{2(\dot{R}^2+kc^2)}{R^2} \geq 0 .
\end{eqnarray}
It is obvious that the functions (\ref{3.7}) and (\ref{3.8})
determine a parametric dependence $\vartheta=\vartheta(\varrho)$,
which in some cases can be reduced to an explicit form by
eliminating $t$.

It should be noted that a similar method was proposed in
\cite{Chervon1,Chervon2}, where the idea that the shape of the
potential of a self-interacting scalar field in standard
inflationary models is not fixed, allows one to obtain new exact
solutions for inflation (in this context see also
\cite{Ellis,Barrow}). In \cite{Chervon1} this approach was called
the method of ``fine turning of the potential''.

\section{Examples}

\subsection{Exact solutions for  the dust model}
To illustrate the above-stated method, we shall give a series of
simple examples. We begin with a dust model ($p=0$, or the
constant $\nu=0$). Let us consider a power-law behaviour of the
scale factor:
\begin{eqnarray} \label{4.1}
R(t)= A t^{n}  \quad ( 1/3 \le n \le 2/3 ),
\end{eqnarray}
where $A$ and $n$ are positive constants and the limits for $n$
were obtained from (\ref{3.9}) and (\ref{3.10}). It is easy to
show that (\ref{4.1}) satisfies (\ref{3.9}) and (\ref{3.10}) for
$k=0$  always and for $k=\pm 1$  only for a restricted period of
time. Further, let us for simplicity restrict our consideration to
the spatially-flat model ($k=0$). With account of (\ref{4.1}) \eqs
(\ref{3.6})--(\ref{3.8}) allow us to find an exact solutions for
$\sigma(t)$, $\dot{\sigma}(t)$, $\varrho(t)$ and $\vartheta(t)$:
\begin{eqnarray} \label{4.2}
\sigma(t)= \pm \; 2\ \sqrt{(2n - 3n^{2})/3}\; \ln{t} + \sigma_{0}  \;,
\end{eqnarray}
\begin{eqnarray} \label{4.3}
\dot{\sigma}(t)= \pm \; 2\;\sqrt{(2n - 3n^{2})/3}\ \frac{1}{t} \;,
\end{eqnarray}
\begin{eqnarray} \label{4.4}
\varrho(t)= \frac{2(3n^{2}-n)}{\kappa_0 c^4} \; \frac{1}{t^{2}} \;,
\end{eqnarray}
\begin{eqnarray} \label{4.5}
\vartheta(t)= \mp \; \frac{(3n-1)\sqrt{3(2n-3n^{2})}}{\kappa_0 c^2} \;
\frac{1}{t^{2}} \;.
\end{eqnarray}
It is interesting to note that the case $n=1/3$ corresponds to a
universe with the scalar $\sigma$-field only (for this case the
exact solutions were found earlier in \cite{Blinkouski1}). The
case $n=2/3$ corresponds to the standard Einstein-de Sitter model
of Friedmann's cosmology. \eqs (\ref{4.4}) and (\ref{4.5}) allow
one to obtain the equation of state $\vartheta=\vartheta(\varrho)$
in an explicit form:
\begin{eqnarray} \label{4.6}
\vartheta= \mp \; \frac{\sqrt{3(2n-3n^{2})}}{2n}\; \varrho c^{2} \quad
(n\not=1/3) \;,
\end{eqnarray}
and also to find the following range for $\vartheta$:
\begin{eqnarray} \label{4.6add}
0 \le \left| \vartheta \right|< 3/2 \varrho c^{2} .
\end{eqnarray}

Let us note the simplest consequences of this model for
observational cosmology. The Hubble parameter and  the age of the
Universe are given by
\begin{eqnarray} \label{4.6a}
H(t)\equiv \frac{\dot{R}}{R}=\frac{n}{t}\; ; \qquad
t_{0}=  \frac{n}{H_{0}}\;, \quad
\frac{1}{3H_{0}}\leq t_{0} \leq \frac{2}{3H_{0}} \;,
\end{eqnarray}
where the subscript 0 denotes the present values. For the
deceleration parameter $q_{0}$ we have
\begin{eqnarray} \label{4.6b}
q_{0}=q \equiv - \frac{\ddot{R}R}{\dot{R}^2}=\frac{1-n}{n} \;,\quad
1/2 \leq q_{0} \leq 2 \;.
\end{eqnarray}
The parameter $\lambda_0$ (already introduced in
\cite{Blinkouski2}), characterizing the $\sigma$-field, is  given
by
\begin{eqnarray} \label{4.6c}
\lambda_0 \equiv \frac{1}{H_0} \frac{\dd\sigma(t_0)}{\dd{t}}
= \pm\; 2\;\sqrt{\frac{1}{n}\left(\frac{2}{3}-n\right)}\;,
\quad 0 \leq \left|\lambda_{0}\right| \leq 2 \; .
\end{eqnarray}
It should be noted that the parameter $\lambda_0$ is, in
principle, a measurable quantity \cite{Blinkouski2}. Also, for the
flat model considered here, $\lambda_0$ and $q_{0}$ are not
independent parameters: they are related by the identity $2q_0 -
3{\lambda_0}^{2}/4  = 1$, while the density parameter is given by
$ \Omega_{0}~=~1-~{\lambda_0}^{2}/4$ and $\Omega_{0}<1$ (for more
detail see \cite{Blinkouski2,Blinkouski3}).

\subsection{Exponential type inflation}
Consider a cosmological model in which  the contribution of vacuum
energy prevails in the total energy density, so that  the equation
of state $p= - \varrho c^{2} $ is realized (see, e.g.,
\cite{Linde}). In this case we suppose that the scale factor
$R(t)$ increases very fast according to the exponential law as in
the classical inflation:
\begin{eqnarray} \label{4.7}
R(t)= A e^{Ht} ,
\end{eqnarray}
where $A$ and $H$ are positive constants. Moreover, here $H$ plays
the role of the  Hubble constant since $\dot{R}/R=H$ for any time.
The expression (\ref{4.7}) satisfies \eqs (\ref{3.9}) and
(\ref{3.10}) for $k=0$ and $k=1$, but they cannot be satisfied if
$k=-1$.

If $k=0$, a substitution of (\ref{4.7}) into
(\ref{3.6})--(\ref{3.8}) gives the simple solution
\begin{eqnarray} \label{4.8}
\sigma=\mbox{const} \; ,\quad \vartheta= 0 \; ,\quad
\varrho=\frac{3H^{2}}{\kappa_0 c^4}=\mbox{const} \;,
\end{eqnarray}
which coincides with the classical de Sitter solution of general
relativity.

If $k=1$, then from (\ref{3.6}) to (\ref{3.8}) we find:
\begin{eqnarray} \label{4.9}
\sigma(t)= \pm \; \frac{2c}{\sqrt{3}AH} \; e^{-Ht} + \sigma_{0}
\;,
\end{eqnarray}
\begin{eqnarray} \label{4.10}
\dot{\sigma}(t)= \mp \; \frac{2c}{\sqrt{3}A} \; e^{-Ht} ,
\end{eqnarray}
\begin{eqnarray} \label{4.11}
\varrho(t)= \frac{3H^{2}}{\kappa_0 c^4} \left(1 + \frac{2
c^2}{3A^{2}H^{2}}\; e^{-2Ht}\right) ,
\end{eqnarray}
\begin{eqnarray} \label{4.12}
\vartheta(t)= \mp \; \frac{2{\sqrt{3}}\ H}{\kappa_0 c A} \; e^{-Ht} \;.
\end{eqnarray}
Notice that these solutions asymptotically tend to (\ref{4.8})
when $t$ is large. Consequently, with an exponential growth of the
scale factor, the $\sigma$-field effects become more and more
negligible, so that at large $t$ the model passes over to the
standard inflationary de-Sitter model. It is interesting to note
that within this simplest inflationary model we get a natural
explanation of why at present we do not observe scalar field
effects or why these effects are so small.

Let us indicate another exponential type of solutions possessing
such properties. Such a solution corresponding to expansion
without a singularity is given by
\begin{eqnarray} \label{4.13}
R(t)= A \cosh{\omega t} ,
\end{eqnarray}
\begin{eqnarray} \label{4.14}
\sigma(t)= \pm \; \frac{2a}{\sqrt{3} \omega} \arctan{(\sinh{\omega t})} +
\sigma_{0} \;,
\end{eqnarray}
\begin{eqnarray} \label{4.15}
\dot{\sigma}(t)= \pm \; \frac{2a}{\sqrt{3}} \; \frac{1}{\cosh{\omega t}} \;,
\end{eqnarray}
\begin{eqnarray} \label{4.16}
\varrho(t)= \frac{3\omega^{2}}{\kappa_0 c^4} \left(1 +
\frac{2a^{2}}{3\omega^{2}}\; \frac{1}{\cosh^{2}{\omega t}} \right) \;,
\end{eqnarray}
\begin{eqnarray} \label{4.17}
\vartheta(t)= \mp \; \frac{2{\sqrt{3}} a \omega}{\kappa_0 c^2} \;
\frac{\sinh{\omega t}}{\cosh^{2}{\omega t}} \;,
\end{eqnarray}
where $a\equiv \sqrt{kc^2A^{-2}-\omega^2}$, $A$ and $\omega$ are
positive constants. In this case the conditions (\ref{3.9}) and
(\ref{3.10}) can only be satisfied if $k=1$.

Next, one can find a solution corresponding to expansion from a
singularity:
\begin{eqnarray} \label{4.18}
R(t)= A \sinh{\omega t} ,
\end{eqnarray}
\begin{eqnarray} \label{4.19}
\sigma(t)= \pm \; \frac{a}{\sqrt{3} \omega} \ln{\frac{\cosh{\omega
t}-1}{\cosh{\omega t}+1}} + \sigma_{0} \;,
\end{eqnarray}
\begin{eqnarray} \label{4.20}
\dot{\sigma}(t)= \pm \; \frac{2a}{\sqrt{3}} \; \frac{1}{\sinh{\omega t}} \;,
\end{eqnarray}
\begin{eqnarray} \label{4.21}
\varrho(t)= \frac{3\omega^{2}}{\kappa_0 c^4} \left(1 +
\frac{2a^{2}}{3\omega^{2}}\; \frac{1}{\sinh^{2}{\omega t}} \right) \;,
\end{eqnarray}
\begin{eqnarray} \label{4.22}
\vartheta(t)= \mp \; \frac{2{\sqrt{3}} a \omega}{\kappa_0 c^2} \;
\frac{\cosh{\omega t}}{\sinh^{2}{\omega t}} \;,
\end{eqnarray}
where now $a\equiv \sqrt{kc^2A^{-2}+\omega^2}$. In this case
(\ref{3.9}) and (\ref{3.10}) is automatically true for $k=1$ and
$k=0$ and can always be satisfied if $k=-1$. It should be noted
that there is one more solution, corresponding to harmonic
behaviour of the scale factor:
\begin{eqnarray} \label{4.23}
R(t)= A \sin{\omega t} ,
\end{eqnarray}
\begin{eqnarray} \label{4.24}
\sigma(t)= \pm \; \frac{a}{\sqrt{3} \omega} \ln{\frac{\cos{\omega
t}-1}{\cos{\omega t}+1}} + \sigma_{0} \;,
\end{eqnarray}
\begin{eqnarray} \label{4.25}
\dot{\sigma}(t)= \pm \; \frac{2a}{\sqrt{3}} \; \frac{1}{\sin{\omega t}} \;,
\end{eqnarray}
\begin{eqnarray} \label{4.26}
\varrho(t)= \frac{3\omega^{2}}{\kappa_0 c^4} \left(-1 +
\frac{2a^{2}}{3\omega^{2}}\; \frac{1}{\sin^{2}{\omega t}} \right) \;,
\end{eqnarray}
\begin{eqnarray} \label{4.27}
\vartheta(t)= \mp \; \frac{2{\sqrt{3}} a \omega}{\kappa_0 c^2} \;
\frac{\cos{\omega t}}{\sin^{2}{\omega t}} \;.
\end{eqnarray}
This is the so-called \cite{Barrow} trigonometric counterpart to
the solution (\ref{4.18}) to (\ref{4.22}).

\subsection{Power law inflation}
As a further example, consider a power-law behaviour of the scale
factor:
\begin{eqnarray} \label{5.1}
R(t)= A t^m \quad ( m \geq 1 ) ,
\end{eqnarray}
where $A$ and $m$ are positive constants. The relation (\ref{5.1})
always satisfies (\ref{3.9}) and (\ref{3.10}) for $k=0$ and for
$k=1$ and can be satisfied for a restricted period of time if
$k=-1$.

If $k=0$, \eqs (\ref{3.6}) to (\ref{3.8}) give the simple solution
\begin{eqnarray} \label{5.2}
\sigma(t)= \pm \; 2\ \sqrt{\frac{m}{3}}\; \ln{t} + \sigma_{0}  \;,
\end{eqnarray}
\begin{eqnarray} \label{5.3}
\dot{\sigma}(t)= \pm \; 2\ \sqrt{\frac{m}{3}}\ \frac{1}{t} \;,
\end{eqnarray}
\begin{eqnarray} \label{5.4}
\varrho(t)= \frac{3m^{2}-m}{\kappa_0 c^4} \; \frac{1}{t^{2}} \;,
\end{eqnarray}
\begin{eqnarray} \label{5.5}
\vartheta(t)= \mp \; \frac{\sqrt{3m}(3m-1)}{\kappa_0 c^2} \; \frac{1}{t^{2}}
\;.
\end{eqnarray}
In this case the equation of state $\vartheta=\vartheta(\varrho)$
in an explicit form is given by
\begin{eqnarray} \label{5.6}
\vartheta= \mp \; \sqrt{3/m}\; \varrho c^{2} \;.
\end{eqnarray}
If $k=\pm 1$, then we have ($m\neq1$)
\begin{eqnarray} \label{5.7}
\sigma(t)= \pm \;\frac{\sqrt{m}}{3(1-m)}\ \left[2 \sqrt{1+at^{-2m+2}} +
\ln{\frac{\sqrt{1+at^{-2m+2}}-1}{\sqrt{1+at^{-2m+2}}+1}} \right] + \sigma_{0}
\;,
\end{eqnarray}
\begin{eqnarray} \label{5.8}
\dot{\sigma}(t)= \pm \; 2\ \sqrt{\frac{m}{3}}\;
\frac{\sqrt{1+at^{-2m+2}}}{t}\;,
\end{eqnarray}
\begin{eqnarray} \label{5.9}
\varrho(t)= \frac{1}{\kappa_0 c^4} \; \frac{3m^2-m+2mat^{-2m+2}}{t^{2}} \;,
\end{eqnarray}
\begin{eqnarray} \label{5.10}
\vartheta(t)= \mp \; \frac{\sqrt{3m}}{\kappa_0 c^2} \;
\frac{3m-1+2mat^{-2m+2}}{t^{2}\; \sqrt{1+at^{-2m+2}}} \;,
\end{eqnarray}
where now $a\equiv kc^2m^{-1}A^{-2}$. It is easy to see that this
solution tends asymptotically to (\ref{5.2})--(\ref{5.5}) at large
$t$. Let us also note that this family of solutions is like the
standard non-inflationary big-bang if $1/3 \le m < 1$. In the case
$m=1$ we have linear inflation. The corresponding solution is
given by
\begin{eqnarray} \label{5.11}
R(t)= A t ,
\end{eqnarray}
\begin{eqnarray} \label{5.12}
\sigma(t)= \pm \; 2\ \sqrt{(1+a)/3}\; \ln{t} + \sigma_{0}  \;,
\end{eqnarray}
\begin{eqnarray} \label{5.13}
\dot{\sigma}(t)= \pm \; 2\ \sqrt{(1+a)/3}\; \frac{1}{t} \;,
\end{eqnarray}
\begin{eqnarray} \label{5.14}
\varrho(t)= \frac{2(1+a)}{\kappa_0 c^4} \; \frac{1}{t^{2}} \;,
\end{eqnarray}
\begin{eqnarray} \label{5.15}
\vartheta(t)= \mp \; \frac{2\ \sqrt{3(1+a)}}{\kappa_0 c^2} \; \frac{1}{t^{2}}
\;,
\end{eqnarray}
where now $a\equiv kc^2A^{-2}$.

\section{Conclusions}
In the present paper, using the arbitrariness in the choice of an
equation of state for the scalaric substrate energy density
$\vartheta = \vartheta(\varrho)$ which is not determined within
the PUFT, we have proposed an exact solutions generation method
for homogeneous and isotropic models of the Universe. This method
has allowed us to find new Friedmann-like solutions for the dust
model as well as solutions for the simplest inflationary  models.
It is interesting to note that within the framework of
exponential-type of inflation we have obtained a natural
explanation of why at present we do not observe $\sigma$-field
effects or why these effects are so negligible. It should be noted
that a cosmological model, in which the $\sigma$-field plays an
essential role both at early stages of the Universe's evolution
and at present, was considered in detail in recent papers by E.
Schmutzer \cite{Schmutzer2000}.

\section*{Acknowledgements}

The authors would like to thank Prof. Ernst Schmutzer for helpful
discussions.


\end{document}